%% file: paper.tex
\documentclass[conference]{IEEEtran}
\IEEEoverridecommandlockouts

\usepackage{cite}
\usepackage{amsmath,amssymb,amsfonts}
\usepackage{algorithmic}
\usepackage{graphicx}
\usepackage{textcomp}
\usepackage{xcolor}
\usepackage{url}
\usepackage{booktabs}
\usepackage{subfigure}
\usepackage[pagebackref,breaklinks,colorlinks]{hyperref}

\def\BibTeX{{\rm B\kern-.05em{\sc i\kern-.025em b}\kern-.08em
    T\kern-.1667em\lower.7ex\hbox{E}\kern-.125emX}}
\begin{document}

\title{
AdaptVC: High Quality Voice Conversion with Adaptive Learning
}

\author{\begin{tabular}{c}
\textit{Jaehun Kim$^1$, Ji-Hoon Kim$^1$, Yeunju Choi$^2$, Tan Dat Nguyen$^1$, Seongkyu Mun$^2$, Joon Son Chung$^1$}\\
$^1$Korea Advanced Institute of Science and Technology, South Korea,
$^2$Samsung Research, South Korea \\
\{kjaehun, jh.kim, tandat.kaist, joonson\}@kaist.ac.kr, 
\{wkadldppdy, skmoon777\}@gmail.com
\end{tabular}
}

\maketitle

\begin{abstract}
The goal of voice conversion is to transform the speech of a source speaker to sound like that of a reference speaker while preserving the original content.
A key challenge is to extract disentangled linguistic content from the source and voice style from the reference.
While existing approaches leverage various methods to isolate the two, a generalization still requires further attention, especially for robustness in zero-shot scenarios.
In this paper, we achieve successful disentanglement of content and speaker features by tuning self-supervised speech features with adapters.
The adapters are trained to dynamically encode nuanced features from rich self-supervised features, and the decoder fuses them to produce speech that accurately resembles the reference with minimal loss of content.
Moreover, we leverage a conditional flow matching decoder with cross-attention speaker conditioning to further boost the synthesis quality and efficiency.
Subjective and objective evaluations in a zero-shot scenario demonstrate that the proposed method outperforms existing models in speech quality and similarity to the reference speech.
\end{abstract}

\begin{IEEEkeywords}
self-supervised learning, speech synthesis, voice conversion
\end{IEEEkeywords}

\input{sections/1_introduction}
\input{sections/2_method}

\input{sections/3_experiment}
\input{sections/4_results}
\input{sections/5_conclusion}

\bibliographystyle{IEEEtran}
\bibliography{IEEEabrv,shortstrings,references}
\end{document}

%% file: sections/1_introduction.tex
\section{Introduction}
Voice Conversion (VC) converts a source speaker's voice to sound as if it were uttered by a target speaker, preserving the original linguistic content.
A powerful voice conversion framework has potential applications, including personalized text-to-speech, privacy security, and language learning tools~\cite{yourtts, voicemixer, crosslingual}.
The quest to closely resemble the target speaker's timbre without losing the original content calls for a successful disentanglement of linguistic and speaker attributes, as well as the generation of rich acoustic representation by effectively fusing the two.
Such a capability has much stronger influence in a zero-shot VC scenario, where the source and target voices are completely unseen during training.

Early VC systems have attempted to resolve disentanglement with various techniques. AutoVC~\cite{qian2019autovc} constructs an autoencoder architecture with an information bottleneck layer to encode content features only, and F0-AutoVC~\cite{f0autovc} extends the idea and utilizes the fundamental frequency to improve the quality of generation.
DiffVC~\cite{diffvc} adopts diffusion mechanism into VC and proposes maximum likelihood sampling that generalizes one-shot VC scenario.
However, the models either fail to generate speech with close resemblance to the reference speaker or lose content information and naturalness.

\begin{figure*}[t]
    \centering
    \includegraphics[width=0.99\textwidth]{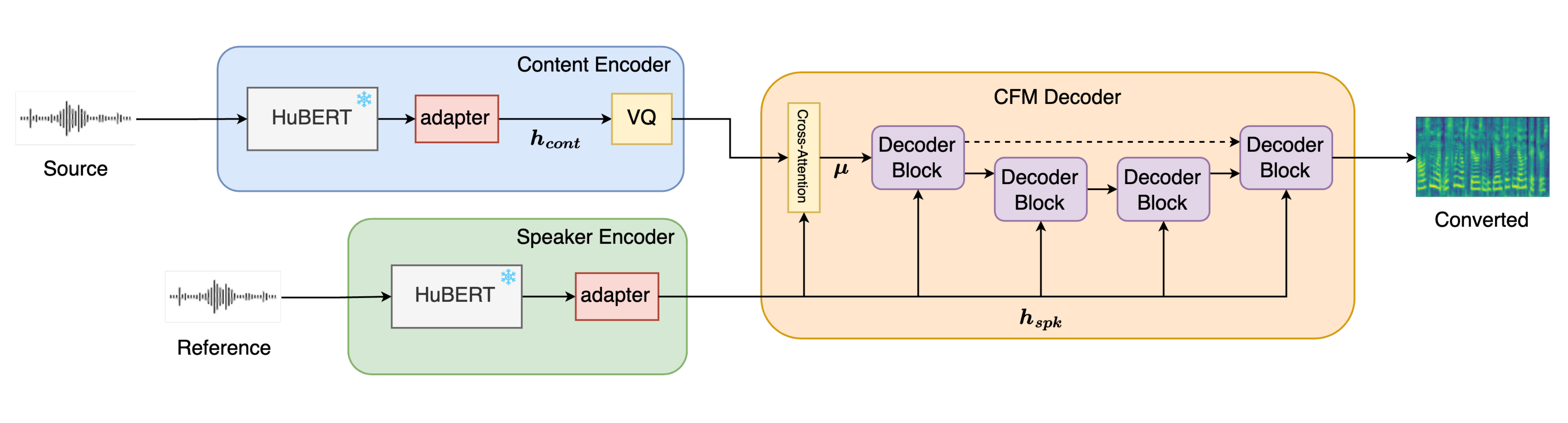}
    \caption{
        Overall architecture of AdaptVC. $\boldsymbol{h_{cont}}$ denotes the content representation from the adapter in the content encoder, and $\boldsymbol{h_{spk}}$ denotes the speaker features from that in the speaker encoder. Prior distribution $\boldsymbol{\mu}$ is obtained by fusing the content and speaker information through cross-attention.
    }
    \label{fig:overall}    
\end{figure*}

Recently, self-supervised learning (SSL) has drawn increased attention for the utilization of large-scale unlabeled data~\cite{hubert, w2v2, wavlm, babu2021xls}.
Moreover, the features extracted from a pretrained SSL model show a high correlation with both the acoustic and linguistic information~\cite{choi2021neural}, suggesting high potential for application in VC research.
NANSY~\cite{choi2021neural} utilizes intermediate features extracted with XLS-R~\cite{babu2021xls} as a content representation disentangled from speaker attributes.
DDDM-VC~\cite{choi2023dddm} follows the similar approach and proposes a dual-path diffusion decoder that separately models source and filter information.
kNN-VC~\cite{knnvc} proposes a non-parametric approach and replaces the features extracted from source speech with WavLM~\cite{wavlm} with the nearest neighbors of those from target speech.
Current SSL-based VC systems have brought the conversion quality close to human, but meticulous parameter searches, including heuristic selection of intermediate layers, and the requirement for large computing time remains unsolved.

To address this, this paper presents AdaptVC, a high-quality voice conversion model with nuanced self-supervised speech representations.
Inspired by the concept of tuning large-scale pretrained models with the small addition of parameters~\cite{pet, lora, adapterhub}, the model incorporates adapters that tunes the rich representation from an SSL model and generates nuanced features.
Specifically, all intermediate layer outputs of an SSL model are combined via weighted summation, and auxiliary modules based on specific objectives automatically guides the model to produce richer representation than a single layer output.
Moreover, high speech quality and fast processing time are achieved through a Conditional Flow Matching decoder with an Optimal Transport objective (OT-CFM)~\cite{cfm} and cross-attention speaker conditioning.
The design allows the decoder to effectively model detailed speaker characteristics by offering multiple conditioning operation.
Both subjective and objective metrics in a challenging zero-shot scenario demonstrate that AdaptVC surpasses all existing voice conversion models in terms of intelligibility and target speaker similarity.
Audio samples are available in the demo page: \url{https://mm.kaist.ac.kr/projects/AdaptVC}

%% file: sections/2_method.tex
\section{Method}
AdaptVC exhibits an encoder-decoder architecture, as illustrated in Fig.~\ref{fig:overall}.
Source and reference utterances are fed to separate encoders comprising HuBERT~\cite{hubert}, a pretrained speech SSL model, with an adapter to combine all intermediate layer outputs.
As illustrated in Fig.~\ref{fig:adapter}, adapters in the encoders contain learnable weights that serve as coefficients for the weighted summation, and the values are updated to maximize the extraction of content and speaker-only information.
The encoded content features are passed to a U-Net based CFM decoder, conditioned with encoded speaker features, which generates the  mel-spectrogram of the converted speech.

\subsection{Content Encoder}
The content encoder aims to extract linguistic features and minimize the influence of speaker-specific attributes.
However, as the objective of training with non-parallel VC is to reconstruct the original speech, the content encoder naturally tends to produce features rich in both content and speaker information.
To further guide the model in disentangling the speaker aspect, a vector quantization (VQ) layer is applied after the adapter.
As demonstrated in \cite{vqgan, vqvae}, the quantization of latent features produces discrete while compact representation.
From the perspective of speech encoding, the output of the adapter is guided to map similar content information from various speakers into closest embedding, ultimately generating accurate linguistic information independent of speakers.

\subsection{Speaker Encoder}
The objective of the speaker encoder is to produce rich speaker features independent of linguistic content.
Unlike conventional approaches where a single vector of speaker information is utilized as a condition, the model leverages frame-wise speaker features to capture the time-varying timbre of different utterances, as \cite{nansy++, elf} demonstrate speech synthesis with close similarity to the target speaker.
A reference speech utterance is passed to the HuBERT model, and the final representation produced by the adapter is fed to the decoder to transform content only features into rich acoustic features.

\begin{figure}
    \centering
    \subfigure[Adapter mechanism]{
        \includegraphics[width=0.4\columnwidth]{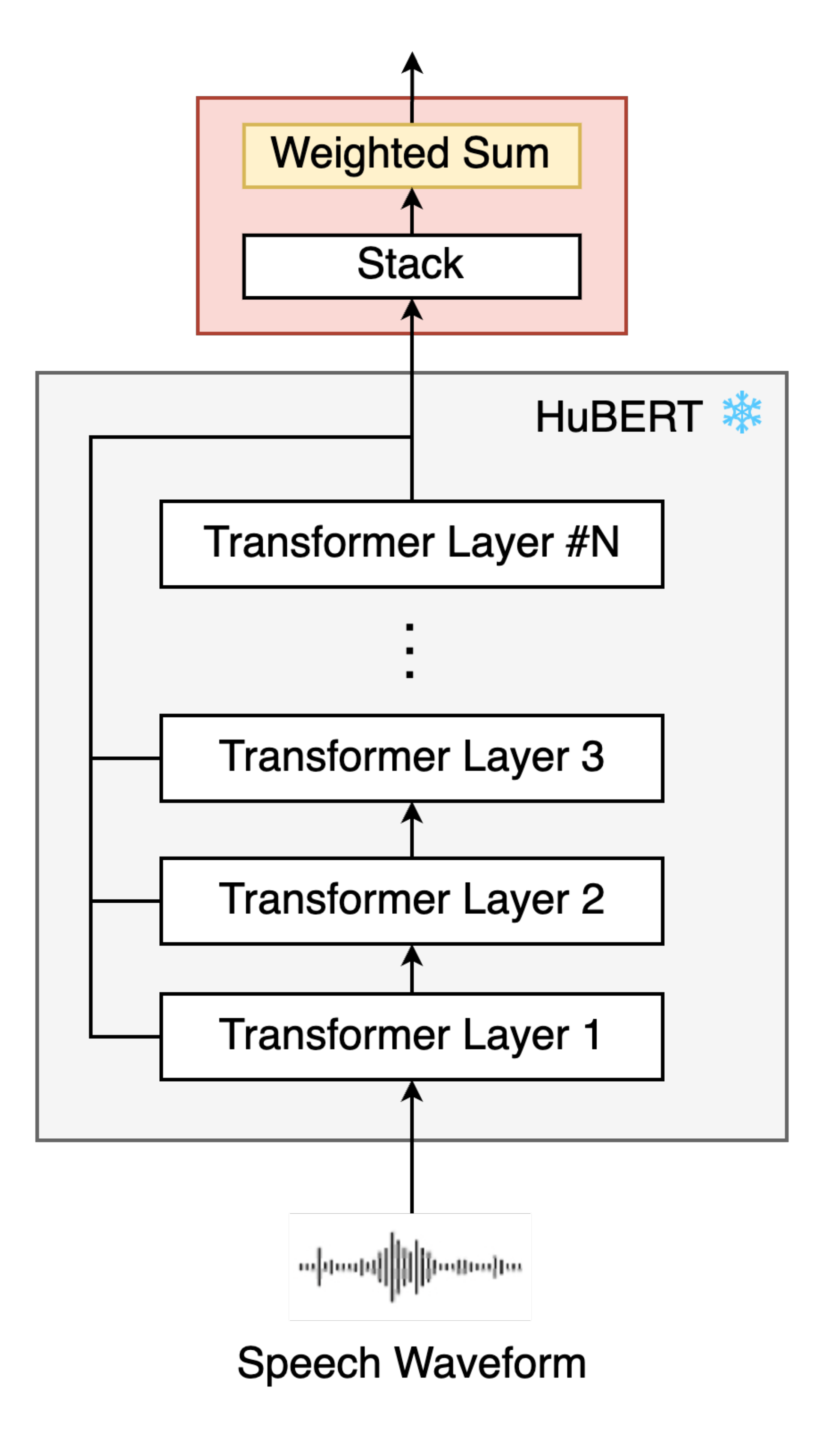}
        \label{fig:adapter}
    }
    \subfigure[Decoder Block]{
        \raisebox{10pt}{
            \includegraphics[width=0.4\columnwidth]{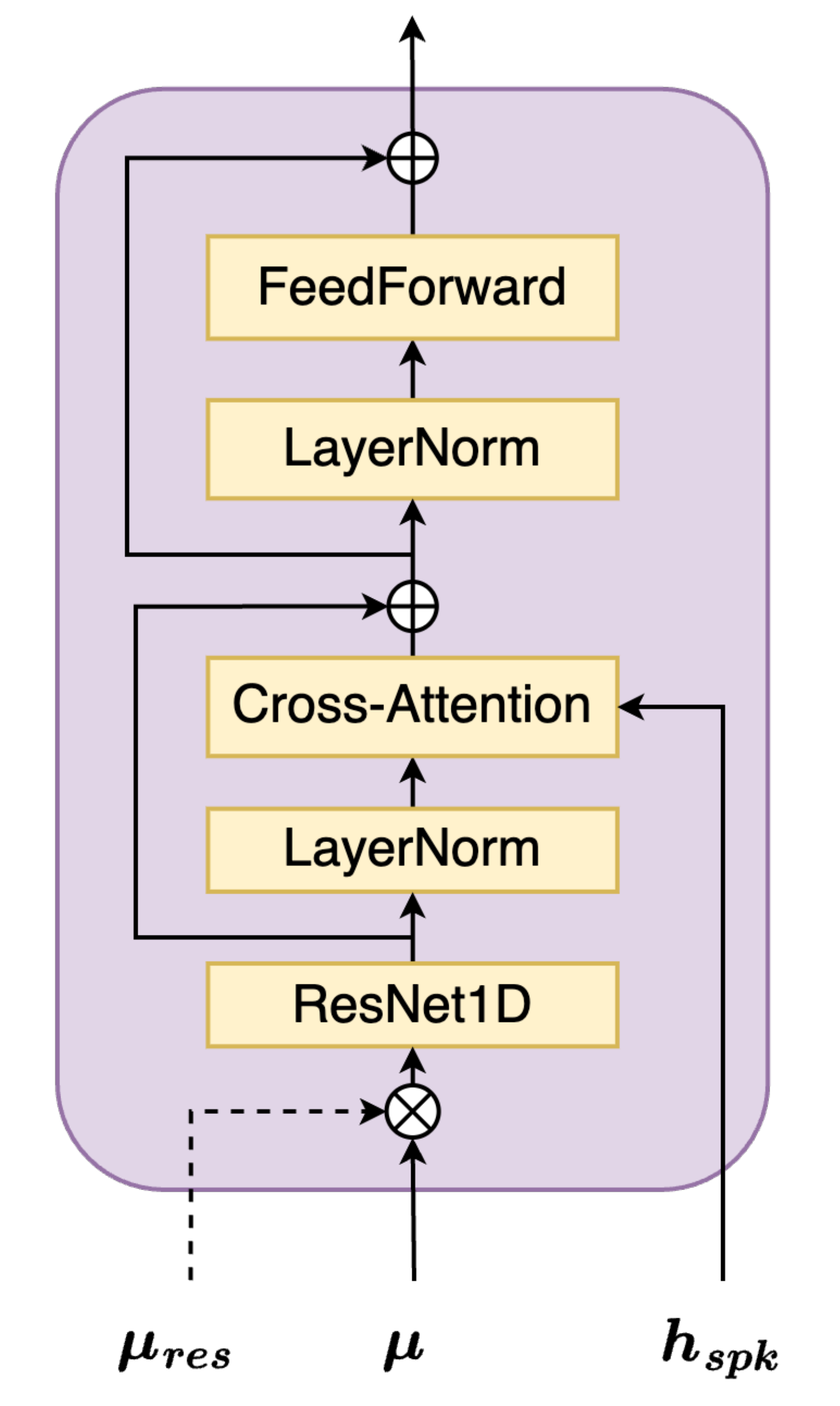}
        }
        \label{fig:decoder}
    }
    \caption{
        Illustration of HuBERT adapter mechanism (a) and decoder block in the CFM decoder (b). Residual input $\boldsymbol{{\mu_{res}}}$ is concatenated to $\boldsymbol{\mu}$ only for blocks with skip connections.
    }
\end{figure}

\subsection{CFM Decoder}
The decoder receives content and speaker features and generates a mel-spectrogram of the converted speech.
To capture efficiency and quality of generation, the model leverages OT-CFM objective~\cite{cfm, mehta2024matcha}.
By regressing the transformation to match the mapping between the data and target distribution, OT-CFM provides higher efficiency and robustness compared to a diffusion mechanism, which models a stochastic transformation of data.
The decoder is designed to provide speaker condition in multiple ways based on a transformer-based U-Net architecture~\cite{mehta2024matcha}.
Inspired by an effective conditioning method in the image domain~\cite{cross-attention}, self-attention layers in the transformer blocks are replaced with cross-attention layers, where the encoded speaker features serve as keys and values as shown in Fig.~\ref{fig:decoder}.
The provision of multiple conditioning via cross-attention allows the decoder to faithfully model the acoustic details of various speakers, and the adapter in the speaker encoder is optimized to produce rich speaker information by combining multiple outputs of HuBERT.

\begin{table*}[t]
    \caption{
        Qualitative and quantitative results with 95\% confidence intervals ($\uparrow$: higher is better, $\downarrow$: lower is better). \\
        The parenthesized numbers indicate the number of sampling steps.
    }
    \label{tab:results}
    \centering
    \resizebox{\textwidth}{!}{
        \begin{tabular}{lccccccc}
        \toprule
        \multicolumn{1}{c}{\textbf{Model}} &
        \multicolumn{1}{c}{\textbf{RTF$\downarrow$}} &
        \multicolumn{1}{c}{\textbf{UTMOS$\uparrow$}} &
        \multicolumn{1}{c}{\textbf{MOS-N$\uparrow$}} &
        \multicolumn{1}{c}{\textbf{MOS-S$\uparrow$}} & 
        \multicolumn{1}{c}{\textbf{WER$\downarrow$}} & 
        \multicolumn{1}{c}{\textbf{CER$\downarrow$}} &
        \multicolumn{1}{c}{\textbf{SECS$\uparrow$}} \\
        
        \midrule
        GT (vocoded)                        & N/A       & $4.07$    &   $4.51\pm0.12$     & $4.04\pm0.21$   & $2.42\pm0.05$ & $1.08\pm0.01$ & $0.982$ \\
        \midrule

        kNN-VC~\cite{knnvc}                 & $0.15$    & $2.87$    & $1.96\pm0.15$    & $2.34\pm0.27$     & $34.6\pm7.53$ &   $21.8\pm3.97$ & $0.752$ \\

        Diff-VC (30)~\cite{diffvc}          & $0.35$    & $3.67$    & $3.33\pm0.14$    & $3.14\pm0.20$      & $13.1\pm1.51$ &   $6.75\pm0.88$    & \textbf{0.828} \\

        Diff-VC (10)~\cite{diffvc}          & $0.22$    & $3.70$    & $3.14\pm0.16$    & $2.92\pm0.20$      & $12.2\pm1.49$ &   $6.22\pm0.87$    & $0.779$ \\
        
        DDDM-VC (30)~\cite{choi2023dddm}    & $0.30$    & $3.43$    & $3.48\pm0.15$    & $3.27\pm0.22$      & $8.32\pm2.09$ &   $4.43\pm1.18$    & $0.819$ \\

        DDDM-VC (10)~\cite{choi2023dddm}    & $0.19$    & $3.51$    & $3.48\pm0.14$    & $3.19\pm0.23$      & $6.40\pm2.15$ &   $3.37\pm1.15$    & $0.823$ \\

        \textbf{AdaptVC (10)}                 & $0.04$    & \textbf{3.95}    & \textbf{4.13} $\pm$ \textbf{0.14}    & \textbf{3.52} $\pm$ \textbf{0.21}      & $7.39\pm1.06$  & $3.63\pm0.58$    & $0.821$ \\

        \textbf{AdaptVC (5)}                  & $0.02$    & 3.94    & $3.86\pm0.14$    & $3.36\pm0.21$     & $6.96\pm0.97$  & $3.29\pm0.49$    & $0.801$ \\

        \textbf{AdaptVC (1)}               & \textbf{0.01}    & $3.38$    & $1.76\pm0.11$    & $2.14\pm0.24$     & \textbf{6.36} $\pm$ \textbf{0.85} & \textbf{2.98} $\pm$ \textbf{0.42}    & $0.768$ \\

        \bottomrule
        \end{tabular}
    }
\end{table*}

\subsection{Training Objective}
The model is trained with three objective functions: commitment loss, prior loss, and OT-CFM loss for the decoder.
The commitment loss enforces the input of the VQ layer to commit to the codebook vectors, as in \cite{vqvae}.
The loss is formulated as:
\begin{equation}
        \mathcal{L}_{commit} = MSE(\boldsymbol{h_{cont}}, \mathrm{sg}[\boldsymbol{e}]),
\end{equation}
where $MSE$ is the mean squared error, $\mathrm{sg}[\cdot]$ is a stop-gradient operator, $\boldsymbol{h_{cont}}$ is the output of the adapter in the content encoder, and $\boldsymbol{e}$ is the codebook vectors in the VQ layer, which become the content features fed to the decoder.

Following \cite{gradtts}, the prior loss minimizes the log-likelihood between the prior distribution and the mel-spectrogram: 
\begin{equation}
    \mathcal{L}_{prior} = -\sum_{i=1}^{T}{\log{\varphi(\boldsymbol{x}_i;\boldsymbol{\mu}_{i}, I)}},
\end{equation}
where $\boldsymbol{x}$ denotes the target mel-spectrogram, $\varphi(\cdot;\boldsymbol{\mu}_{i},I)$ is a probability density function of $\mathcal{N}(\boldsymbol{\mu}_{i}, I)$, and $T$ denotes the temporal length.
The prior loss also directs the codebook vectors in the VQ layer to represent discrete but nuanced information.

The loss for the decoder follows \cite{mehta2024matcha}, which estimates a vector field with linear trajectory via optimal transport (OT):
\begin{equation}
\begin{split}
\mathcal{L}_{dec} = \mathbb{E}_{t, q(\boldsymbol{x}_1), p_0(\boldsymbol{x}_0)}\Vert{u}^{\text{OT}}_t(\phi^\text{OT}_t(x_0) | \boldsymbol{x}_1) \\-{v}_t(\phi^{\text{OT}}_t(x_0) | \boldsymbol{\mu}, \boldsymbol{{h}_{spk}}; \theta) \Vert^2,
\end{split}
\end{equation}
where $\theta$ denotes the network parameters, $\phi^{OT}_t(\boldsymbol{x}) = (1 - (1 - \sigma_{min})t)\boldsymbol{x}_0 + t \boldsymbol{x}_1$ represents flow that maps the source and target distribution, $u_t$ is a known vector field that generates approximate path from prior distribution $p_0$ to target data distribution $p_t$.
$\boldsymbol{h_{spk}}$ represents continuous speaker features obtained from the speaker encoder, serving as a condition.

Finally, the total training objective is formulated as:
\begin{equation}
    \mathcal{L}_{total} = \mathcal{L}_{commit} + \mathcal{L}_{prior} + \mathcal{L}_{dec}.
\end{equation}

%% file: sections/3_experiment.tex
\section{Experiment}

\subsection{Data}
The model was trained with LibriTTS~\cite{libritts}, a multi-speaker text-to-speech dataset, where train-clean-100 and train-clean-360 subsets were split into training, validation, and test sets following~\cite{choi2023dddm}.
To evaluate zero-shot VC performance, 20 source speakers and 20 target speakers from the VCTK~\cite{vctk} corpus were utilized.
Speech waveforms were resampled to 16kHz to match the input configuration of HuBERT.
A log-scale mel-spectrogram was generated with window size and filter size of 1280, hop size of 320, and mel filterbank size of 80, where the temporal resolution was also adjusted to match that of HuBERT.

\subsection{Training}
The two adapters in the content and speaker encoders were constructed with a single fully-connected layer without bias, followed by a softmax activation to ensure the sum of the probabilities is 1.
The VQ layer in the content encoder employed a single quantizer with codebook size of 512.
The architecture of the decoder followed that of \cite{mehta2024matcha}, where self-attention layers in each decoder block were replaced with cross-attention.
A HiFi-GAN~\cite{hifigan} vocoder was trained with the training subset to correctly evaluate the performance of the proposed model.

\subsection{Baselines}
The performance of the model was compared with three established VC models: kNN-VC~\cite{knnvc}\footnote{\url{https://github.com/bshall/knn-vc}}, DDDM-VC~\cite{choi2023dddm}\footnote{\url{https://github.com/hayeong0/DDDM-VC}}, DiffVC~\cite{diffvc}\footnote{\url{https://github.com/trinhtuanvubk/Diff-VC}}.
The samples were generated based on the official implementations.
Since the reported kNN-VC model is trained with a different dataset, we trained the model with LibriTTS to ensure a fair comparison.
The samples were normalized with respect to the root mean square value before evaluation.

\subsection{Evaluation Metrics}
Both qualitative and quantitative metrics were obtained for comprehensive evaluation.
Mean Opinion Score (MOS) was conducted to 20 domain experts with 40 generated samples.
Naturalness (MOS-N) was evaluated from the perspective of general speech quality and the intelligibility by presenting the ground-truth text for a reference, and perceptual similarity to the reference speaker (MOS-S) was measured by juxtaposing the parallel target speech.
Moreover, an automatic speech quality estimation model (UTMOS)~\cite{utmos} was utilized.
Word Error Rate (WER) and Character Error Rate (CER) were calculated by a pretrained ASR model~\cite{whisper}.
Speaker Embedding Cosine similarity (SECS) between the generated speech and the target ground-truth is obtained by calculating cosine similarity of the speaker embeddings obtained through \texttt{Resemblyzer}\footnote{\url{https://github.com/resemble-ai/Resemblyzer}}, a pretrained speaker verification model~\cite{wan2018generalized}.
Lastly, Real Time Factor (RTF) was measured for speed comparison.

%% file: sections/4_results.tex
\section{Results}

\subsection{Quantitative Evaluation}
The quantitative metrics (WER, CER, RTF) in Table~\ref{tab:results} demonstrate that the proposed method consistently achieves high intelligibility.
The proposed model with a single sampling step results in the lowest WER and CER and RTF.
Although increasing the number of sampling steps leads to higher error rates, the observed differences are smaller than those in the baseline models, which highlights the robustness of the proposed method.
In conclusion, the proposed model with 5 sampling steps shows the best balance between performance and speed.

\subsection{Qualitative Evaluation}
As shown in Table~\ref{tab:results}, the proposed method outperforms existing approaches in naturalness and similarity MOS by a significant margin.
DiffVC achieves the highest SECS due to its direct use of speaker vectors from the same model~\cite{wan2018generalized}, while the proposed approach attains the highest perceptual similarity, as reflected in the high MOS-S values.
Moreover, AdapterVC with only 5 sampling steps clearly outperforms the baseline models with its speed up to 10 times faster.
This demonstrates the strong performance of the proposed method as well as its applicability to real-time scenarios.

\begin{figure}[ht]
    \centering
    \includegraphics[width=\columnwidth]{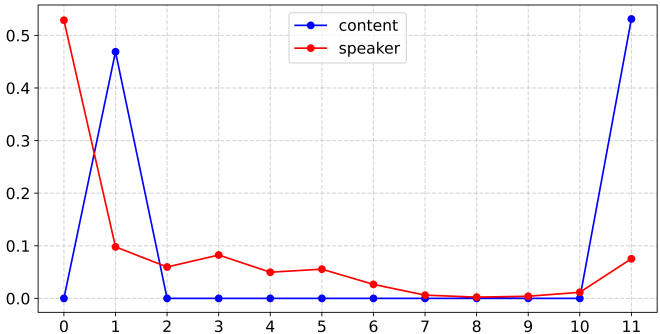}
    \caption{Visualization of adapter weights. Numbers in the x-axis indicate layer indices and y-axis denotes the trained weights.}
    \label{fig:afe}
\end{figure}

\subsection{Analysis on Adapter Weights}
The weights of the adapters in content and speaker encoders exhibit distinct behaviors, as visualized in Fig.~\ref{fig:afe}.
The adapter in the content encoder mainly utilizes the second and last layer output of HuBERT, while the weights of the other layers converge to zero.
The adapter in the speaker encoder, on the other hand, shows the highest weight on the first layer and gradually decreases as the layers proceed.
This aligns with the findings in \cite{weightedsum}, where a speech SSL model shows a tendency to capture acoustic information in the earlier layers and robust linguistic information in the latter, and consequently validates the adaptability of the proposed method.

\begin{table}[th]
  \caption{Qualitative (UTMOS) and quantitative (WER, SECS) results of ablation study.}
  \centering
  \label{tab:ablation}
  \resizebox{0.99\columnwidth}{!}{
      \begin{tabular}{lcccc}
        \toprule
        \multicolumn{1}{c}{} &
        \multicolumn{1}{c}{\textbf{UTMOS}} &
        \multicolumn{1}{c}{\textbf{WER$\downarrow$}} & 
        \multicolumn{1}{c}{\textbf{SECS$\uparrow$}} \\
        
        \midrule
        \textbf{AdaptVC (5)}              & 3.94    & 6.96 $\pm$ 0.97 & 0.801 \\
        ~~\textit{w/o adapters}           & 3.81    & 8.47 $\pm$ 0.90 & 0.793 \\
        ~~\textit{w/o VQ}                 & 3.90    & 1.52 $\pm$ 0.36 & 0.648 \\
        \midrule
        ~~\textit{condition: SALN}        & 3.76    & 12.4 $\pm$ 1.73 & 0.754 \\
        ~~\textit{condition: Mean + Add}  & 3.76    & 12.8 $\pm$ 1.90 & 0.756 \\
        \bottomrule
      \end{tabular}
      }
\end{table}

\subsection{Ablation Study}
To validate the contribution of each module in the proposed method, a systematic ablation study was performed.
First, the adapters were replaced with the fixed output -- the last layer output for the content encoder, and the first layer for the speaker encoder, as they show high correlation to the linguistic and speaker information~\cite{ssl-layer2}.
The impact of the VQ layer was evaluated by removing it.
The contribution of the cross-attention speaker condition was compared with two established speaker conditioning methods: Style Adaptive Layer Normalization (SALN)~\cite{metastylespeech} and mean pooling followed by its addition to latent content features.

The removal of adapters show a clear decline in the generated speech's naturalness and intelligibility, as shown in Table~\ref{tab:ablation}.
This indicates that the learned combination of multiple intermediate outputs from HuBERT contains more nuanced information compared to a single layer output.
While the model without the VQ layer shows notably high UTMOS and low WER, the similarity of the converted speech to the reference speaker is significantly low.
Qualitatively, the model merely reconstructs the source speech regardless of the reference speaker, underscoring the VQ layer's importance in disentangling content.
Finally, a distinct degradation in speaker similarity with the two conditioning methods provides clear evidence that speaker conditioning with multiple cross-attention contributes to accurately resemble the reference speaker.

%% file: sections/5_conclusion.tex
\section{Conclusion}

This paper proposes AdaptVC, a high quality voice conversion model with adaptive learning framework.
The proposed adapters automatically determine the optimal combination of intermediate SSL layer outputs, thereby eliminating the need for heuristic parameter tuning and the integration of additional information.
The utilization of OT-CFM decoder and speaker conditioning with multiple cross-attention layers efficiently boost the quality of generation.
Qualitative and quantitative evaluations suggest that AdaptVC outperforms the existing approaches by a significant margin.